\documentclass[epj]{svjour}
\usepackage{graphicx}
\begin{document}
\title{Perfect transfer of quantum states in a network of harmonic oscillators}
\author{D. Portes Jr\inst{1}\thanks{\email{portes@hotmail.com}} \and H. Rodrigues\inst{1}\thanks{\email{harg@cefet-rj.br}} \and S. B. Duarte\inst{2}\thanks{\email{sbd@cbpf.br}} \and B. Baseia\inst{3}\thanks{\email{baseiabasilio@gmail.com}} 
}                     
%
%
\institute{Centro Federal de Educa\c{c}\~ao Tecnol\'ogica do Rio de Janeiro, 
Av. Maracan\~a 229, 20.271-110, Rio de Janeiro, RJ, Brazil \and Centro Brasileiro de Pesquisas F{\'i}sicas, 
Rua Dr. Xavier Sigaud 150, 22.290-180, Rio de Janeiro, RJ, Brazil \and Instituto de F{\'i}sica, Universidade Federal de Goi\'as, 
PO-BOX-131, 74.001-970, Goiania, GO, Brazil}
\date{Received: date / Revised version: date}
%
\abstract{
 This work presents an exactly soluble scheme to address
the problem of optimal transfer of quantum
states through a set of $s$ harmonic oscillators
composing a  network with connected ends as a closed quantum circuit. For
this purpose we start from a general quadratic
Hamiltonian form. The relationship between the
parameters of the Hamiltonian, the network size,
and the time interval required for such transfer
are explicitly shown. Particular physical realizations of this Hamiltonian, transfer
 of entangled states, including transfer of states at the
 expense of a quantum entanglement, are also considered.
}%
%
\maketitle
\section{Introduction}
\label{intro}

Among the various interesting problems studied in quantum optics one may
cite (and distinguish) the teleportation of states between two (non
interacting) quantum systems \cite{Bennett}, from one subsystem to the
other, and alternatively, the transfer of states between (interacting)
quantum systems \cite{Jahne,Portes}. In the first case the process
occurs crucially due to the intervention of an entangled state that
describes a combined bipartite system. In the second case the process is due
to an appropriate type of interaction between the subsystems. This second
scenario may also include the study of exchange of states between the
subsystems \cite{Dodonov,Rodrigues,Bastos,Mollow}. In both scenarios the efficiency
of the process deserves special attention, whose verification involves the
calculation of the fidelity and the success probability of the operation.
The case of state transfer is also interesting, e.g., in the realm of
quantum spin networks \cite{Matthias,Franco,Jafarizadeh}. Others have
investigated the behavior of a quantum state propagating through a network
of interacting oscillators \cite{Plenio,Audenaert}, with few
attention on how to find out the conditions that govern the connected
oscillators to allow transmission of states closer to the ideal case. The
issue concerns the perfect transfer of states in terms of fidelity and
success probability, including the transfer of entangled states. 

 In this work we pursue the answer to the following
query: what is, if any, the appropriated class of Hamiltonian that allows us to get such
a state transfer through the coupled oscillators network? To this end, we developed a compact method to obtain a
Hamiltonian form that produces a perfect state transfer through the
mentioned network. In particular, we explicitly determine this 
Hamiltonian form for a time independent $s$-sized network with connected
ends. Such network configuration simulates a closed quantum circuit of
oscillators which is important to discuss possible occurrence of
nonclassical effects, as state revival, squeezing, and others during the
state propagation. We show that, for a perfect quantum state transfer
along the system, the coupling can not be restricted to next neighbors, but
it must embrace the whole network. Transfer of correlated quantum states in the network was also studied, including 
examples where these type of states play an auxiliary role for state transfers.

This paper is organized as follows. In the Section \ref{sec:2} we develop the mathematical
basis for our analysis to describe the time evolution of the characteristic
function of the state of our $s$-sized HO-network. In the Section \ref{sec:3} we use the
result of Section \ref{sec:2} in order to obtain a suitable Hamiltonian form that yields perfect
cyclic transfer of an arbitrary state. In the Section \ref{sec:4} we discuss the time evolution of a quantum (number) state and also the
transfer of entangled states. The Section \ref{sec:5} treats the propagation of the (nearest classical) coherent 
state and discusses the physical realizations of our employed Hamiltonian form, including the transfer of states at the expense of a quantum entanglement. The Section \ref{sec:6} contains the comments and conclusions.

\section{The characteristic function of the system state}
\label{sec:2}

In the Schr\"{o}dinger picture, the density operator $\mathbf{\rho} (t)$ for the state of $s$ identical coupled
harmonic oscillator may be described by means of the characteristic function $%
\chi$, in the form \cite{Glauber}
\begin{equation}
\mathbf{\rho}(t)=\pi^{-s}\int \chi \left( \alpha _{1}, \ldots ,\alpha_{s};t \right)^s_{j=1} \left[ \mathbf{D}_{j}^{-1}\left( \alpha_{j}\right) d^{2}\alpha _{j}\right] ,  \label{1.1}
\end{equation}%
with $\chi $\ defined in terms of $s\ $complex quantities $\alpha _{j}$ as,%
\begin{equation}
\chi \left( \alpha _{1}, \ldots ,\alpha _{s};t\right) \equiv Tr\left[ \mathbf{%
\rho }(t)^s_{j=1}\mathbf{D}_{j}\left( \alpha _{j}\right) %
\right] ,  \label{1.2}
\end{equation}%
where
\begin{equation}
\mathbf{D}_{j}\left(\alpha \right) \equiv \exp \left( \alpha \mathbf{a}%
_{j}^{\dag }-\alpha ^{\ast }\mathbf{a}_{j}\right) .  \label{1.3}
\end{equation}%
The integration in Eq. (\ref{1.1}) is carried out in the whole $\alpha _{j}$-plane, 
and the element $d^{2}\alpha _{j}$ is defined by
\begin{equation}
d^{2}\alpha_{j} = d\left({Re}(\alpha _{j})\right) d\left({Im}(\alpha
_{j})\right) .
\end{equation}

The reduced density matrix $\rho _{k}$ for a $k$-th subsystem is obtained
from the partial trace of $\rho (t),$ taken over all other subsystems,%
\begin{equation}
\mathbf{\rho }_{k}(t)=Tr_{1}\left[ ...Tr_{j\neq k}\left[ ...Tr_{s}[\mathbf{%
\rho }(t)]\right] \right] .  \label{1.4}
\end{equation}%
By considering that 
\begin{equation}
\mathbf{\rho }_{k}(t)=\frac{1}{\pi }\int \chi _{k}(\alpha ;t)\mathbf{D}%
_{k}^{-1}\left( \alpha \right) d^{2}\alpha ,  \label{1.6}
\end{equation}%
and using Eqs. (\ref{1.1}) and (\ref{1.4}) with the properties of the characteristic function
from,
\begin{equation}
Tr[\mathbf{D}(\alpha )]=\pi \delta ^{(2)}(\alpha ),  \label{1.7}
\end{equation}%
the reduced characteristic function can be written as, 
\begin{equation}
\chi _{k}(\alpha ;t)\equiv \left. \chi \left( 0,...,\alpha
_{k},...0;t\right) \right\vert _{\alpha _{k}=\alpha } .  \label{1.5}
\end{equation}

A complete transfer of the quantum states of the $m-th$ oscillator to the $%
n-th$ one is obtained by imposing the prescription,
\begin{equation}
\chi _{n}(\alpha ;\tau )=\chi _{m}(\alpha ;0),  \label{1.8}
\end{equation}%
or%
\begin{equation}
\chi \left( 0,...,\alpha _{n},...0;\tau \right) =\chi \left( 0,...,\alpha
_{m},...0;0\right) .  \label{1.9}
\end{equation}%
The prescription (\ref{1.8}) indicates that the state transfer is realized after the elapsed time $\tau$.

Next, we use the well known functional identity for the characteristic
function of a system composed by $s$ quantum-mechanical oscillators \cite%
{Mollow},
\begin{equation}
\chi \left( \alpha _{1},...,\alpha _{s};t\right) =\chi (\alpha
_{1}(t),\ldots ,\alpha _{s}(t);0),  \label{1.10}
\end{equation}%
where the complex quantities $\alpha _{j}(t)$\ are obtained from the inverse
of the Bogoliubov transformation of the Heisenberg operators, 
\begin{equation}
\mathbf{a}(t)=\mu (t)\mathbf{a}(0)+\nu (t)\mathbf{a}^{\dag }(0),
\label{1.11}
\end{equation}%
namely, 
\begin{equation}
\alpha (t)=\mu ^{\dag }(t)\alpha (0)-\nu ^{T}(t)\alpha ^{\ast }(0) ,
\label{1.12}
\end{equation}%
where the matrix representation was used for operators and coefficients in (\ref{1.11})-(\ref{1.12}), ${\mathbf a}=[a_{j}]_{s\times 1},$ $\alpha =[\alpha
_{j}]_{s\times 1},$ $\mu =[\mu _{jk}]_{s\times s}$ and $\nu =[\nu
_{jk}]_{s\times s}$. It should be emphasized that the relation (\ref {1.12})
 is valid only for the Bogoliubov transformation, when the Hamiltonian is
quadratic in the time-independent creation and annihilation operators, $\mathbf{a}$
and $\mathbf{a}^{\dag}$. From Eqs. (\ref{1.5}), (\ref{1.10}) and (\ref{1.12}) the
temporal evolution of the reduced characteristic function is given by
\begin{equation}
\chi _{n}\left( \alpha ;t\right) =\chi (\mu _{n1}^{\ast }(t)\alpha -\nu
_{n1}(t)\alpha ^{\ast },\ldots,\mu _{ns}^{\ast }(t)\alpha -\nu
_{ns}(t)\alpha ^{\ast };0).  \label{1.13}
\end{equation}%
From Eq. (\ref{1.8}), we have the condition,
\begin{equation}
\left\{ 
\begin{array}{c}
\mu _{nj}^{\ast }(\tau )\alpha -\nu _{nj}(\tau )\alpha ^{\ast }=\alpha
~~~\left( j=m\right) \\ 
\mu _{nj}^{\ast }(\tau )\alpha -\nu _{nj}(\tau )\alpha ^{\ast }=0~~~\left(
j\neq m\right)%
\end{array}%
\right. ,  \label{1.14}
\end{equation}%
for arbitrary $\alpha$. The solution of equation (\ref{1.14}) is%
\begin{equation}
\left\{ 
\begin{array}{l}
\mu_{nj}(\tau) = \delta_{jn}, ~~~ j=1,\ldots,s \\ 
\nu_{nj}(\tau) = 0, ~~~~~~ j=1,\ldots,s%
\end{array}%
\right.  . \label{1.15}
\end{equation}

Particularly, for a cyclic permutation representing the state transfer of an
oscillator to one of its first neighbor, and considering the boundary
condition of coincident network ends, plus the conditions 
\begin{equation}
\chi _{m}(\alpha ;\tau )=\chi _{m-1}(\alpha ;0) \; {\rm  and } \; \chi _{1}(\alpha
;\tau )=\chi _{s}(\alpha ;0),  \label{1.16}
\end{equation}%
we have%
\begin{equation}
\mu (\tau )=C\equiv \left( 
\begin{array}{ccccc}
0 & 0 & \cdots & 0 & 1 \\ 
1 & \ddots & \ddots & \ddots & 0 \\ 
0 & \ddots & \ddots & \ddots & \vdots \\ 
\vdots & \ddots & \ddots & \ddots & 0 \\ 
0 & \cdots & 0 & 1 & 0%
\end{array}%
\right)  \label{1.17}
\end{equation}%
and%
\begin{equation}
\nu (\tau )=0 .  \label{1.18}
\end{equation}%
Substituting Eqs. (\ref{1.17}), (\ref{1.18}) and (\ref{1.12}) in Eq. (\ref%
{1.10}), we find that%
\begin{equation}
\chi \left( \alpha _{1},...,\alpha _{s};\tau \right) =\chi (\alpha
_{2},...,\alpha _{s},\alpha _{1};0).  \label{1.19}
\end{equation}%
Thus, any state transfer between HO-oscillators in the network corresponds
to a cyclic permutation of the arguments in the characteristic
function of the whole system after a time interval $\tau$. This result can
be generalized to any type of permutation, hence not restricted to a cyclic
one.

\section{The Hamiltonian for perfect state transfer}
\label{sec:3}

To construct the time-independent Hamiltonian that describes the dynamics
illustrated in the previous section, we consider a general quadratic form in
the creation $\mathbf{a}^{\dag}$ and annihilation $\mathbf{a}$ operators 
\begin{equation}
\mathbf{H}=\hbar \sum_{j,k=1}^{s}~\lambda _{jk}\mathbf{a}_{j}^{\dag }\mathbf{%
a}_{k}+\hbar \sum_{j,k=1}^{s}\left( \gamma _{jk}\mathbf{a}_{j}^{\dag }%
\mathbf{a}_{k}^{\dag }+\gamma _{jk}^{\ast }\mathbf{a}_{j}\mathbf{a}%
_{k}\right) .  \label{2.1}
\end{equation}%
This Hamiltonian can be diagonalized by a Bogoliubov transformation given by 
\begin{equation}
\mathbf{a}_{k}^{\prime}=\sum_{j=1}^{s}\left( W_{kj}\mathbf{a}_{j}+V_{kj}%
\mathbf{a}_{j}^{\dag }\right) ,  \label{2.2}
\end{equation}%
which leads it to the Hamiltonian form 
\begin{equation}
\mathbf{H}=\hbar \sum_{j}^{s}\omega _{j}~\mathbf{a}_{j}^{\prime \dag }%
\mathbf{a}_{j}^{\prime }.  \label{2.2a}
\end{equation}

By taking the time evolution of the Heisenberg operators $\mathbf{a}(t)$ in
Eq. (\ref{1.11}) we obtain,
\begin{eqnarray}
\mu (t) &=&W^{\dag }e^{-i\Omega t}W-V^{T}e^{i\Omega t}V^{\ast } ,  \label{2.3}
\\
\nu (t) &=&W^{\dag }e^{-i\Omega t}V-V^{T}e^{i\Omega t}W^{\ast } ,
\label{2.4}
\end{eqnarray}%
where 
\begin{equation}
\Omega =diag\left( \omega _{1},...,\omega _{s}\right)  .  \label{2.5}
\end{equation}%
To have the complete transfer of state, as mentioned in previous section
(see Eqs. (\ref{1.17}) and (\ref{1.18})), the matrices should satisfy $\mu (\tau
)=C$ and $\nu (\tau )=0,$ thus we have%
\begin{equation}
W^{\dag }e^{-i\Omega \tau }W-V^{T}e^{i\Omega \tau }V^{\ast }=C  , \label{2.6}
\end{equation}%
and%
\begin{equation}
W^{\dag }e^{-i\Omega \tau }V-V^{T}e^{i\Omega \tau }W^{\ast }=0 .
\label{2.7}
\end{equation}%
A proof of uniqueness of Bogoliubov transformation satisfying the last two
equations is shown in Appendix A. This transformation is defined by $V=0$ and 
the unitary matrix $W$ that diagonalizes $C$, given explicitly by \cite{Iachello},
\begin{equation}
W_{jk}=\frac{1}{\sqrt{s}}e^{2\pi ijk/s} .  \label{2.9}
\end{equation}%
Thus, we have 
\begin{equation}
WCW^{\dag }=diag(e^{-2\pi i\left( 1/s\right) },e^{-2\pi i\left( 2/s\right)
}...,e^{-2\pi i}) .  \label{2.10}
\end{equation}%
From Eq. (\ref{2.6}), and taking $V=0$, we obtain 
\begin{equation}
WCW^{\dag }=e^{-i\Omega t} .  \label{2.8}
\end{equation}%
Equations (\ref{2.10}) and (\ref{2.8}) lead to the relation%
\begin{equation}
\omega _{j}=\frac{2\pi }{\tau }\left( \frac{j}{s}+m_{j}\right)  ,
\label{2.11}
\end{equation}%
where $m_{j}$\ are arbitrary integers. We will impose that all $m_{j}\geq 0$
to ensure positive eigenvalues for the Hamiltonian. For $m_{j}=0$ and any $%
j$ we obtain the frequencies of the fundamental modes.

As $V=0$ we see from Eq. (\ref{2.2}) that the terms $\gamma_{jk} {\mathbf a}_{j}^{\dag
}{\mathbf a}_{k}^{\dag }+\gamma _{jk}^{\ast }{\mathbf a}_{j} {\mathbf a}_{k}$ are excluded in the
Hamiltonian (\ref{2.1}). So the general form of our Hamiltonian is given by%
\begin{equation}
\mathbf{H}=\hbar ~\mathbf{a}^{\dag }W^{\dag }\Omega W\mathbf{a} .
\label{2.12}
\end{equation}%
 This Hamiltonian form propagates an arbitrary state of an oscillator in
the network, allowing its complete transfer to another one after the time
interval $\tau$. More explicitly, substituting the form of $\Omega $\ from
Eq. (\ref{2.5}) and that of $W$ from Eq. (\ref{2.9}) in Eq. (\ref{2.12}) 
we obtain the coupling coefficients of the original Hamiltonian (\ref{2.1})%
\begin{equation}
\lambda _{jk}=\frac{2\pi \hbar }{s\tau }\sum_{l=1}^{s}\left( \frac{l}{s}%
+m_{l}\right) \exp \left( 2\pi i\left( j-k\right) \frac{l}{s}\right) .  \label{2.13}
\end{equation}%
Since we have arbitrary integers $m_{l}$ in the last equation, we indeed
obtained an infinite and numerable family of Hamiltonian forms (labeled by
the $m_{l}$-values), allowing the perfect state transfer. From Eq. (\ref%
{2.13}) (with $k=j$) we can see that the diagonal elements, 
\begin{equation}
\lambda _{jj}=\frac{2\pi \hbar }{s\tau }\sum_{l=1}^{s}\left( \frac{l}{s}%
+m_{l}\right) ,  \label{2.14}
\end{equation}%
are directly related with the trace of $\Omega$ matrix. Note that, in
principle, none of the coupling $\lambda_{jk}$ are null. So, during the
elapsed time $\tau$ the perfect transfer of states between oscillators is
an effect of collective coupling between oscillators in the network.

Next, substituting $V=0$\ into Eqs. (\ref{2.3}) and (\ref{2.4}), one gets%
\begin{eqnarray}
\mu (t) &=&W^{\dag }e^{-i\Omega t}W  \label{2.19} \\
v(t) &=&0 ,  \label{2.20}
\end{eqnarray}%
and then%
\begin{equation}
\mu _{jk}(t)=\frac{1}{s}\sum_{l=1}^{s}\exp \left( 2\pi i\left[ j-k-\frac{t}{%
\tau }\right] \frac{l}{s}\right) \exp (-i\frac{2\pi }{\tau }m_{l}t) .
\label{2.21}
\end{equation}%
The following properties are satisfied by the these matrices:%
\begin{eqnarray}
\mu _{jk}(t) &=&\mu _{sk}(t-\tau ),\ \; {\rm for \ } j=1,  \label{2.23} \\
\mu _{jk}(t) &=&\mu _{(j-1)k}(t-\tau ),\ \; {\rm for \ } 2\leq j\leq s ,  
\end{eqnarray}%
and%
\begin{eqnarray}
\mu _{js}(t) &=&\mu _{jk}(t-\tau ),\ \; {\rm for \ } k=1,  \label{2.24} \\
\mu _{jk}(t) &=&\mu _{j\left( k+1\right) }(t-\tau ), \; {\rm for\ }2\leq
k\leq s  .  
\end{eqnarray}%
From these properties and knowing that $\mu (0)=1$, we have $\mu (t)$ for
all multiple of the period $\tau$%
\begin{equation}
\mu (n\tau )=\mu (\tau )^{n} .  \label{2.25}
\end{equation}%
In special, for\ $n=s$ we have $\mu (s\tau )=1$, as expected, since we are
using coincident ends for the network, as boundary condition.

\section{Number state and entanglement transfer}
\label{sec:4}

Here we will firstly analyze the case where the initial configuration
represents the first oscillator prepared in a number state $|n\rangle$, while the
others in their fundamental modes, i.e.,%
\begin{equation}
|\Psi (0)\rangle =|n,0,....,0\rangle .  \label{3.1}
\end{equation}%
Choosing the first oscillator instead of any other is irrelevant since the
Hamiltonian is symmetrical by any cyclic permutation of the oscillator
label, as seen in the previous section. The characteristic function
associated with the state at $t=0$ is 
\begin{equation}
\chi (\alpha _{1},...,\alpha _{s};0)=f^{(n)}(\alpha _{1})f^{(0)}(\alpha
_{2})....f^{(0)}(\alpha _{s}) , \label{3.2}
\end{equation}%
where \cite{Glauber}%
\begin{equation}
f^{(n)}(\alpha )=e^{-\frac{1}{2}\alpha ^{\ast }\alpha}L_{n}(\alpha^{\ast}\alpha ) ,
\label{3.3}
\end{equation}%
and $L_{n}(x)$ stands for the Laguerre polynomial. Thus, taking into account that $L_{0}(x)=1$, we get
\begin{equation}
\chi (\alpha _{1},...,\alpha _{s};0)=L_{n}(\alpha _{1}^{\ast }\alpha
_{1})\exp \left( - \frac{1}{2} \sum_{j=1}^{s}\alpha _{j}^{\ast }\alpha _{j}\right) .
\label{3.4}
\end{equation}%
The time evolution of the reduced characteristic function can be easily
obtained by using equation (\ref{1.13}) with $\nu (t)=0$.  We thus find
\begin{equation}
\chi _{j}(\alpha ;t)=e^{-\frac{1}{2}\alpha^{\ast}\alpha}L_{n}\left[ g_{j}(t)\alpha
^{\ast }\alpha \right]  ,  \label{3.5}
\end{equation}%
where 
\begin{equation}
g_{j}(t)=\mu _{j1}^{\ast }(t)\mu _{j1}(t) .  \label{3.6}
\end{equation}%
Note that the sum $\sum \alpha_{j}^{\ast }\alpha_{j}$ is time invariant
due to relation $\mu^{\dag}\mu =1.$ According to equation (\ref{3.5}) the $%
j-th$ oscillator is in its ground state when $g_{j}(t)=0$ and in a number
state when $g_{j}(t)=1.$ From the property given by Eq. (\ref{2.23}) we have
\begin{equation}
g_{j}(t)=g_{1}\left( t-(j-1)\tau \right)  .  \label{3.7}
\end{equation}%
Therefore all oscillators in the network exhibit similar time evolution,
except for a time delay.

\begin{figure}[htpb]
\centering 
\includegraphics[width=.2\textheight]{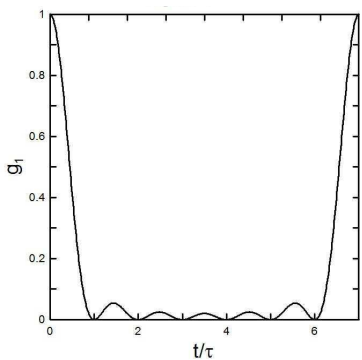}
\vspace*{-.1cm} 
\caption{Function $g_{1}(t),$ defined in the text, with no excitation ( $m_{j}=0$ for all oscillators) in a network
with size $s=7$. \label{fig:100}}
\end{figure}

\begin{figure}[htpb]
\centering 
\includegraphics[width=.2\textheight]{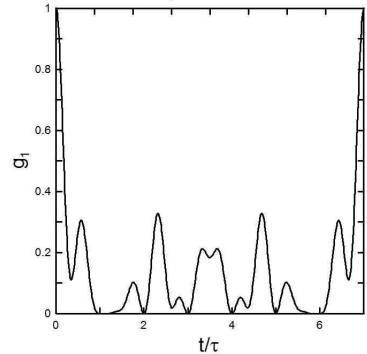}
\vspace*{-.1cm} 
\caption{Function $g_{1}(t),$ defined in the text, for $m_{2}=1,$\ $m_{6}=2$\ and $m_{j}=0$ for other oscillators in a
network with $s=7$. \label{fig:200}}
\end{figure}

Figure \ref{fig:100} shows the function $g_{1}(t)$\ with $m_{j}=0$\ for all $j$, for $s=7
$. We can see that $g(t)$\ has the expected behavior: for values of time
multiples of $\tau$ ($t=k\tau$) we have either $g_{1}=0$\ or $g_{1}=1$. We
can also see that the transferred state runs cyclically through the network
with the\ period $s\tau $. Figure \ref{fig:200} shows the function $g_{1}(t)$\ for $%
m_{2}=1,$\ $m_{6}=2$\ and $m_{j}=0$\ for other $j$. For $t=k\tau$ the
behavior of $g_{1}(t)$, showed in figure \ref{fig:200}, remains the same; however for $t\neq k\tau 
$ this behavior is quite different with additional excitations, which
reflects the higher energy level of the Hamiltonian.

It is also easy seeing that an initial entanglement, in the sense of EPR
state \cite{EPR} will propagate through the network. For example, if we take 
\begin{equation}
|\Psi (0)\rangle =\frac{1}{\sqrt{2}}\left( |n,0\rangle +|0,n\rangle \right)
\times |0,....,0\rangle  ,  \label{3.12}
\end{equation}%
due to the linear time evolution we have for $t=\tau $, 
\begin{eqnarray}
|\Psi (0)\rangle &=&\frac{1}{\sqrt{2}}|0,n,0....,0\rangle +\frac{1}{\sqrt{2}}%
|0,0,n,0....,0\rangle  \label{3.13} \\
&=&\frac{1}{\sqrt{2}}|0\rangle \times \left( |n,0\rangle +|0,n\rangle
\right) \times |0,....,0\rangle .  \label{3.14}
\end{eqnarray}%
The same occurs for any time multiple of $\tau$ and oscillator
interactions that occur during the state propagation restore the
entanglement whenever $t=k\tau$. In addition, if any state correlations is
produced by the oscillator interactions itself, they should vanish
periodically at every time $t=k\tau$.

\section{Coherent state propagation}
\label{sec:5}

Let us now suppose the first oscillator in a coherent state $|\beta\rangle$ at $%
t=0$, again with the others in their fundamental mode, i.e.,%
\begin{equation}
|\Psi (0)\rangle =|\beta ,0,....,0\rangle  ,  \label{3.15}
\end{equation}%
with $\ a_{1}|\Psi (0)\rangle =\beta |\Psi (0)\rangle$. The
characteristic function associated with the coherent state $|\beta\rangle$ is 
\begin{equation}
f^{(\beta )}(\alpha )=e^{-\alpha ^{\ast }\alpha /2}e^{\alpha \beta ^{\ast
}-\alpha ^{\ast }\beta } .  \label{3.16}
\end{equation}%
Thus,%
\begin{equation}
\chi (\alpha _{1},...,\alpha _{s};0)=\exp \left( \alpha _{1}\beta ^{\ast
}-\alpha _{1}^{\ast }\beta \right) \exp \left( \sum_{j=1}^{s}\alpha
_{j}^{\ast }\alpha _{j}\right) .  \label{3.17}
\end{equation}%
Analogously to the result obtained in Eq. (\ref{3.5}), we have%
\begin{equation}
\chi _{j}(\alpha ;t)=\exp \left( -\alpha ^{\ast }\alpha /2+\mu
_{j1}(t)\alpha \beta ^{\ast }-\mu _{j1}^{\ast }(t)\alpha ^{\ast }\beta
\right) ,  \label{3.18}
\end{equation}%
and the propagation of coherent states is such that at $t=k\tau \ $the
coherence is restored.

As physical realizations of the Hamiltonian model in Eq. (\ref{2.1}) we mention that: (i) it describes a set of $s$ noninteracting HO (or equivalently, $s$ field modes) when only $\lambda_{jj}$ are the non null coefficients; (ii) when $\lambda_{jk} = \gamma_{jk} = 0$ for all $j \neq k$ it recovers the quadratic (``two-photon'') Hamiltonian which also responds for the generation of squeezing effect (see Ref.  \cite{Walls}). One origin/realization of the ``two-photon'' Hamiltonian comes from the $\mathbf{x} \mathbf{p} + \mathbf{p} \mathbf{x}$ interaction, where $\mathbf{x} \propto (\mathbf{a} + \mathbf{a}^{\dag})$ and $\mathbf{p} \propto (\mathbf{a} - \mathbf{a}^{\dag})$ stand for the position and momentum operators of the HO (or the quadratures of a field mode), respectively; (iii) when $\gamma_{jk} = 0$ for all $j$ and $k$, it becomes bilinear in operators $\mathbf{a}$ and $\mathbf{a}^{\dag}$ and concerns a set of $s$ interacting HO (or field modes), constituting the so 
called ``one-photon'' interaction. In this case and in the interaction picture it recovers, for $s=2$, the ``beam-splitter''
(BS) Hamiltonian for traveling fields, which furnishes the state of the output field from a given input field that impinges a BS as in \cite{Gerry}. Thus, with $\lambda_{12} = \lambda_{21} = \lambda$ and $ |\Psi \rangle_{\rm in} = |1_{1},0_{2} \rangle $, we get
\begin{eqnarray}
|\Psi\rangle_{\rm out} &=& \exp\left(-i \tau \frac{\mathbf{H}}{\hbar}\right) |\Psi \rangle_{\rm in}  \nonumber \\
&=& \exp [-i \lambda (\mathbf{a}^{\dag}_1 \mathbf{a}_2 + \mathbf{a}_1 \mathbf{a}^{\dag}_2) ] |1_{1},0_{2} \rangle \nonumber \\
& =&  \cos( \lambda \tau) |1_{1},0_{2}\rangle + i \sin(\lambda \tau) |0_{1},1_{2}\rangle, \label{588}
\end{eqnarray}
$\cos( \lambda \tau)$ and $\sin( \lambda \tau)$ being the transmission and reflection coefficients of the BS. Next, if we let the output  (\ref{588}) working as an input for a second BS (named BS$_{2}$ to differ it from the first BS, now named BS$_{1}$), similar calculations result in the new output ($T_i = \cos(\lambda \tau_i)$ and $R_i =\sin(\lambda \tau_i)$ being the transmission and reflection coefficients of the BS$_{i}$, respectively),
\begin{eqnarray}
|\Psi_{2}\rangle_{\rm out} &=& (T_1 T_2-R_1 R_2)|1_{1},0_{2}\rangle \nonumber \\
&+& i (R_1 T_2 + R_2 T_1)|0_{1},1_{2}\rangle , \label{599}
\end{eqnarray}
that reduces to (for $R_1=T_1$ and $R_2=T_2$, or $R_1=T_2$ and $R_2=T_1$),   
\begin{equation}
|\Psi_{2}\rangle_{\rm out} = |0_{1},1_{2}\rangle ,
\end{equation}
which can also be viewed as a perfect transfer: of the  
one-photon state going from the field mode-1 of BS$_{1}$ to the field 
mode-2 of BS$_{2}$. Here, contrary to the classical case where transfer 
occurs between interacting subsystems, the two BS's do not  
interact and the event comes from intervention of the quantum 
entanglement in Eq. (\ref{599}). 

These examples and that of Ref. \cite{Portes} constitute
some applications of the Hamiltonian model (\ref{2.1}). 
Besides the HO's and field modes, applications can be extended to nanotechnological devices
 as  superconducting quantum circuits based on Cooper pair boxes, nanoresonators, etc. \cite{Armour} .

\section{Final remarks and conclusions}
\label{sec:6}

In this work we have studied the state transfer between coupled HO in a $s$-sized network. The
procedure is based on the dynamic evolution of the system described by the
characteristic function introduced in section II. With this description we
have shown that the complete exchange of state between oscillators can be
accomplished in a characteristic elapse of time. We have determined the
explicit Hamiltonian form that allows such state transfer. The procedure is
somewhat similar to that of ``engineering Hamiltonians''. In this way, and
exploring properties of the characteristic function of the system, we have
also analyzed the transfer of a genuine quantum state and also the transfer
of coherent states. For initial entangled states, it was shown that they are
restored whenever the state transfer is accomplished.

Before finalizing, we call attention for the particular case $s = 2$, where
one recovers the results found in Ref. \cite{Portes} using the Wigner approach. In Ref. \cite{Portes} cyclical transfer of states was studied via the Hamiltonian,  
\begin{eqnarray}
\mathbf{H}= \frac{\pi \hbar}{\tau}  && \left[ \left( \frac{3}{2} + m_{1} + m_{2}\right) 
\left( \mathbf{a}_{1}^{\dag }\mathbf{a}_{1}+\mathbf{a}
_{2}^{\dag }\mathbf{a}_{2}\right) \right. \nonumber \\
+ && \;  \left. \left( \frac{1}{2}+m_{2}-m_{1}\right)
\left( \mathbf{a}_{2}^{\dag }\mathbf{a}_{1}+\mathbf{a}_{1}^{\dag }\mathbf{a}
_{2}\right) \right] , \label{2.15}
\end{eqnarray}
which can be put in the more compact form, 
\begin{equation}
\mathbf{H}=\omega \hbar \left( \mathbf{a}_{1}^{\dag }\mathbf{a}_{1}+\mathbf{a%
}_{2}^{\dag }\mathbf{a}_{2}\right) +c\hbar \left( \mathbf{a}_{2}^{\dag }%
\mathbf{a}_{1}+\mathbf{a}_{1}^{\dag }\mathbf{a}_{2}\right) .  \label{2.16}
\end{equation}%
Note that this Hamiltonian provides the relation between the coupling
constant $c$\ and the characteristic field frequency $\omega $%
\begin{equation}
\frac{c}{\omega }=\frac{1+2m_{2}-2m_{1}}{3+2m_{1}+2m_{2}} ,  \label{2.17}
\end{equation}%
and also the characteristic elapse of time $\tau \ $value,%
\begin{equation}
\tau =\left( \frac{1}{2}+m_{2}-m_{1}\right) \frac{\pi }{c} .  \label{2.18}
\end{equation}

Thus, for two oscillators the state transfer occurs when they have the same
frequency, with the coupling parameter satisfying Eq. (\ref{a8}) for
arbitrary non-negative integers $m_{1}$ and $m_{2}$. According to equation (\ref{2.17}), we
always have $c<\omega$. The case $c\ll$ $\omega$ is
usually assumed in quantum optics, and periodical transfer of states occurs
at $t=(m+1/2)\pi /c$, with $m=m_{2}-m_{1}$ being an arbitrary integer,
which recovers our result in Ref. \cite{Portes}.

\section{Acknowledgment}
The authors thank the Brazilian agency CNPq for the partial support.

\section{Appendix A}
\label{sec:7}

All Bogoliubov transformation, represented by the matrices $W$ and $V,$ must
satisfy four conditions \cite{Tikochinsky}:%
\begin{equation}
WW^{\dag }-VV^{\dag }=1 ,  \label{a1}
\end{equation}%
\begin{equation}
W^{\dag }W-V^{T}V^{\ast }=1 ,  \label{a2}
\end{equation}%
\begin{equation}
WV^{T}-VW^{T}=0 ,  \label{a3}
\end{equation}%
and%
\begin{equation}
W^{\dag }V-V^{T}W^{\ast }=0 .  \label{a4}
\end{equation}%
Besides, we need to solve the Eqs. (\ref{2.6}) and (\ref{2.7}), namely%
\begin{equation}
W^{\dag }e^{-i\Omega \tau }W-V^{T}e^{i\Omega \tau }V^{\ast }=C , \label{a5}
\end{equation}%
and%
\begin{equation}
W^{\dag }e^{-i\Omega \tau }V-V^{T}e^{i\Omega \tau }W^{\ast }=0 ,  \label{a6}
\end{equation}%
where $C$ is a unitary matrix. To this end we multiply from the left the Eq. (\ref{a6})
by $W$ and replace Eqs. (\ref{a1}), (\ref{a3}) and (\ref{a5}) to get
\begin{equation}
e^{-i\Omega \tau }V=VC .
\end{equation}
Next, multiplying Eq. (\ref{a5}) from the left by $W$, and replacing the
Eqs. (\ref{a1}), (\ref{a3}) and (\ref{a6}), we obtain
\begin{equation}
e^{-i\Omega \tau }W=WC .  \label{a8}
\end{equation}%
From Eq. (\ref{a1}) and $\left\vert \det (W)\right\vert \geq 1$, then $W$ is
invertible and%
\begin{equation}
e^{-i\Omega \tau }=WCW^{-1} .  \label{a9}
\end{equation}%
The matrix that diagonalizes a unitary matrix is also unitary, thus $%
WW^{\dag }=1$ and $V=0$.

\end{document}